# Electric power transfer in spin pumping experiments

K. Rogdakis[1], N. Alfert[1], A. Srivastava[2], J.W.A. Robinson[2], M. G. Blamire[2], L. F. Cohen[3], and H. Kurebayashi[1,a)]

[1]*London Centre for Nanotechnology, University College London, 17-19 Gordon Street, London, WC1H 0AH, United Kingdom*

[2]*Department of Materials Science and Metallurgy, University of Cambridge, 27 Charles Babbage Road, Cambridge CB3 0FS, United Kingdom*

[3]*The Blackett Laboratory, Imperial College London, SW7 2AZ, United Kingdom*

**a)** Electronic mail: h.kurebayashi@ ucl.ac.uk

**Abstract**

Spin pumping is becoming an established method to generate voltages from magnetic dynamics. The standard detection method of spin pumping is based on open circuit voltage measurement across ferromagnetic (FM) and non-magnetic (NM) bi-layers, where the inverse spin-Hall effect (ISHE) can convert spin currents into electrical charge accumulation. In this paper, we present that it is also possible to measure the associated electric charge current generated in FM/NM bi-layers, by using a macroscopic closed circuitry detection method. Using variable load resistors connected in series to the sample, we quantified charge currents and associated electric power dissipation as a function of the load resistance. By using basic circuit analysis, we are able to describe spin pumping cells as a non-ideal voltage source or equivalent current source with an internal resistor.

## I. INTRODUCTION

Spin currents, the flow of angular momentum without the simultaneous transfer of electrical charge, play a pivotal role in the current developments of spintronic research. For example, they can be used to manipulate magnetic dynamics[1], to switch magnetic moments[2] and to facilitate spintronic energy harvesting applications[3]. The generation of spin currents has been widely explored in the last decade[4]. Through the ISHE[5] charge currents can be converted into a spin current that flows perpendicular to the charge flow orientation and a wide range of thin-film metallic and semiconducting materials have been examined for efficient generation and detection of spin currents through the SHE[6,7,8]. Another widely-used method to generate spin currents is spin-pumping using magnetic dynamics[9]. When magnetic dynamics is driven by microwave (MW) absorption of a FM material, non-equilibrium spin-waves are accumulated and dissipated at a fix rate. One of the dissipation mechanisms in a FM/NM bilayer is to transfer angular momentum to electron spins in the NM layer where spin currents are then generated. The spin currents are then converted into charge currents through the ISHE as illustrated in Fig. 1a. High spin-orbit materials such as Pt are often used as a good SHE material for efficient detection of the spin currents. Although this is an electric *charge-current* excitation picture, the commonly-used circuitry for spin-pumping measurements is based on an open circuit where *electric voltages* are measured and discussed as a signature of spin pumping and the ISHE.

In this paper, we introduce a charge current detection method of spin pumping in a macroscopic closed circuit. In this scheme, instead of using an open circuit, we close the circuit with a load resistor which allows us to measure the generated charge current in a spin-pumping device using the ISHE. In doing so, we are able to measure electric power transfer from the spin pumping device into the external load resistor, which was previously impossible otherwise. Furthermore, our basic circuit analysis indicates that a non-ideal voltage (or equivalent current) source model can describe spin pumping devices. From this analysis, the maximum power transfer, internal resistance as well as ISHE current for the devices were evaluated and discussed. This type of analysis has not been widely employed, only very recently in a spin-Seebeck experiment[10] and not in particular for spin pumping experiments using ferromagnetic resonance (FMR). We also add that a similar charge detection method for measuring the ISHE has been reported by Omori *et al.*[11]. However, in contrast to our study, their devices are a micro-fabricated closed circuit where spin-currents were generated through an electrical spin injection method. Our study demonstrates the wide applicability of the closed-loop circuit measurements for spin pumping as it is not limited to lithographically-patterned nano-scaled devices, but is also applicable to macroscopic spin pumping devices where no micro-fabrication steps are needed.

## II. EXPERIMENTAL

Pt (5 nm)/Ni$_{81}$Fe$_{19}$ (5 nm) bilayers were grown by DC magnetron sputtering onto 5 mm × 5 mm area thermally-oxidised single crystal silicon substrates in Ar at 1.5 Pa in an ultrahigh vacuum system with a base pressure better than $10^{-9}$ mBar. Layer thicknesses were controlled through deposition power and by rotating substrates below stationary magnetrons. Following growth, the samples were placed face-down on a co-planar waveguide board to maximise the oscillating magnetic field strength and an insulating ultra-thin tape was inserted between the sample and the MW board. Two electrodes were attached at both ends of the sample chips allowing electrical detection of the ISHE. For ISHE measurements, we pulse-modulated the MW to increase sensitivity.

## III. RESULTS AND DISCUSSION

We measured FMR MW absorption in our sample by ac magnetic field modulation techniques through a MW power detector. As shown in Fig. 1(b) we observe a clear FMR peak around 30 mT at MW frequency of 4 GHz – note that the differential form of a Lorentzian curve has been measured due to the ac modulation techniques. We then measured the voltage across the sample ($V_{\text{open}}$) while sweeping the same magnetic field region and observed Lorentzian-type peaks as shown in Fig. 1(c). Consistent with the symmetry of electric field generated by the ISHE ($\boldsymbol{E_{ISHE}} \propto \boldsymbol{\sigma} \times \boldsymbol{J_s}$), a sign change has been observed in $V_{\text{ISHE}}$ when the magnetic field polarity is reversed; here, $\boldsymbol{\sigma}$ is the spin polarisation direction and $J_s$ is the spin current flow orientation. Furthermore, we plot the MW power dependence of the peak voltage ($\Delta V_{open}$) in Fig. 1(d) confirming that the dependence follows a standard model of spin-pumping in the linear regime[12]. We note that there might be other voltage components within $\Delta V_{open}$ we quantified, such as spin rectification voltages arising from the planar Hall effect that in theory appears in the Lorentzian lineshape[13,14]. A clear separation between these voltages and the ISHE one is difficult and hence we treat $\Delta V_{open}$ as a voltage generated by spin precession in general.

In order to extract electric power from spin pumping, we used a variable load resistor to close the open circuit configuration as shown in Fig. 2a. By doing this, pumped charge currents in the sample are able to flow in the closed circuit and hence finite power dissipation will take place in the load resistor, which can be measured through the voltage drop across the load resistor. Figure 2b plots the voltage drop ($V_{load}$) for different load resistors. As one can see, the voltage across the resistor is small at low load resistance ($R_{load}$) values and increases by increasing $R_{load}$. By plotting the voltage amplitude ($V_{peak}$) as a function of $R_{load}$ in Fig. 2c, we observe the gradual growth of $V_{peak}$ towards the open circuit voltage ($V_{open}$) in the high resistance regime. We repeated the same experiments for different MW powers, confirming the same dependence. Using this voltage, $V_{peak}$, and the known value of $R_{load}$ for each measurement, we calculated the charge current flowing across the load resistor ($I_{load}$) and plotted it in Fig. 3a. Likewise, the dissipated power across the load resistors ($P_{dis}$) has also been calculated using $P_{dis} = I_{load}V_{load}$ and plotted in Fig. 3b. We observe that $P_{dis}$ peaks at finite $R_{load}$ and tends to zero for both the smallest and highest load resistance limits.

These results from the closed circuit experiments can be understood by a non-ideal voltage source model as shown in Fig. 4a. A non-ideal voltage source includes an internal resistor to self-dissipate power. When the internal resistor is in series to $R_{load}$, the (fixed) pumped voltage has to be distributed into the two resistors, each dissipating electric power. This series resistor circuit acts as a potential divider and $V_{load}$ can be expressed as:

$$V_{load} = \frac{R_{load}}{R_{load} + R_i} V_{open}$$

where $R_i$ is the internal resistance. This equation reproduces the $R_{load}$ dependence of our observed voltages $V_{peak}$ shown as solid curves in Fig. 2c. The charge current and dissipated power in the load resistors can be also calculated by using $R_i$ and $V_{open}$ as:

$$V_{load} = \frac{V_{open}}{R_{load}+R_i}, \quad P_{dis} = \frac{R_{load}}{(R_{load}+R_i)^2} V_{open}^2.$$

$P_{dis}$ has the peak value of $V_{peak}^2/4R_i$ when $R_{load} = R_i$; this is the maximum power transfer condition, frequently observed and discussed in standard electronics such as radio transmitters and high frequency amplifiers[15]. Using the above theoretical formulae we are able to fit our experimental data (solid lines of Fig. 2c, 3a and 3b) with $R_i$ as the only fitting parameter. Despite the wide range of $R_{load}$ (six orders of magnitude), all best-fit results can be produced with $R_i$ very close to the measured sample resistance of 29 Ω (see Table I). From this, we can conclude that there exists an internal resistance in samples during spin pumping experiments, which should be taken into account when one tries to transfer electric power from a spin pumping device to an external load resistor.

TABLE 1. Internal resistance ($R_i$) extracted from simultaneous fits to $V_{peak}$, $I_{load}$ and $P_{dis}$ data for each measurement with MW power ranging from 20 to 200mW.

| MW power [mW] | 20 | 40 | 80 | 100 | 160 | 200 |
|---|---|---|---|---|---|---|
| $R_i$ [Ω] | 38.2 | 28.8 | 40.2 | 36.8 | 29.0 | 38.1 |

We would like to emphasise that we can also consider the spin pumping device cell as a non-ideal current source. A transform of a voltage source to an equivalent current source is possible through Thevenin-Norton circuit analysis[16]. According to Norton's theorem, any network of batteries and resistors having two output terminals can be replaced by the parallel combination of a current source and a resistor. We show in Fig. 4 the two equivalent circuits for describing spin pumping experiments where the current source value can be found by: $I_0 = V_0/R_i$. Using the current source model where the load and internal resistors are in parallel, we can also write $V_{load}$ as $V_{load} = (\frac{R_{load}R_i}{R_{load}+R_i})I_0$. This equation with the current source explains measured voltages such as $V_{peak}$ in Fig. 2c very well - for small $R_{load}$, the prefactor of $\frac{R_{load}R_i}{R_{load}+R_i}$ remains small, meaning that $V_{peak}$ in experiment should be small, whereas $V_{\text{peak}}$ gradually increases with increasing $R_{load}$, eventually approaching to the value of $V_{open}$ for $R_{load} = \infty$ (the open circuit condition).

Finally, we discuss a phenomenological model that takes into account the conversion of MW energy into magnetic energy and then the magnetic energy into electric energy. Using data in Fig. 1(a), the known modulation microwave field amplitude ($h_{\text{mod}}$), the equation: $V_{diode-dc} = (1/h_{mod}) \int V_{lock-in}(H)dH$ (here $V_{diode-dc}$ is the calculated dc response in the diode detector) and the diode voltage-power conversion ratio, we calculated the MW absorption peak for the FMR and estimated the peak height 4 mW for 200 mW measurements. Compared with 14 fW which is the highest electric power measured for the same MW insertion, we have a significant loss (a factor of 3.5×10$^{-12}$) in energy transfer between the MW absorption into electric power generation in the load resistor. This lossy system can be readily understood since magnetic dynamics are *per se* very lossy where a considerable amount of excited angular momentum (and energy) is constantly damped into the lattice. In addition, the spin pumping process extracts a fraction of the non-equilibrium angular momentum into the conduction electron spin system in the adjacent non-magnetic layer, which depends hugely on the efficiency of spin coupling at the interface, i.e. the mixing conductance. A phenomenological model of spin pumping[17,18,19] can indicate that the energy transfer ratio is proportional to $\left(g_{mix}^{\uparrow\downarrow}\right)^2/\alpha^3$, which reflects on the discussions above; here $g_{mix}^{\uparrow\downarrow}$ is the real part of the mixing conductance[20], and $\alpha$ is the Gilbert damping parameter. In other words, layer structures with a better damping material as well as higher mixing conductance will lead to a better energy transfer in spin pumping measurements.

## IV. CONCLUSIONS

In summary, we have shown that it is possible to measure an electric charge current generated by spin-pumping and the ISHE, by using a macroscopic closed circuitry detection method. Using variable load resistors, we are able to quantify electric power generated by spin pumping and successfully transfer magnetic energy into electric energy. We applied basic circuit analysis to our results and found that spin pumping cells can be described by a non-ideal voltage source, or equivalent current source, whose internal resistance is very close to the measured device resistance.

ACKNOWLEDGEMENTS

This work was supported by the EPSRC through Programme Grant EP/N017242/1.

**Figure captions**

FIG. 1: (a) A schematic of spin-pumping and ISHE mechanism in a ferromagnetic (FM) and non-magnetic (NM) bilayer sample. (b) FMR spectra measured at frequency of 4GHz and 20mW MW power using ac-modulation techniques. (c) Voltage ($V_{open}$) measured across the sample for different powers. Results for both positive (solid lines) and negative (dashed lines) magnetic field directions are shown for comparison. The measurement MW frequency is 4 GHz and magnetic field was applied perpendicular to the voltage measurement direction in-plane. (d) MW excitation power dependence of voltage peak height $\Delta V_{open}$ (dot) with a linear fit line (solid curve).

FIG. 2: (a) **A s**chematic of the closed-circuit measurement setup we used for this study. We measured voltage drops across a variable resistor ($V_{load}$) placed in series with the spin-pumping device. Magnetic field was applied perpendicular to the voltage measurement direction in-plane. (b) Measurement of $V_{load}$ for different load resistor values ($R_{load}$). The MW power was 100mW. (c) Peak Amplitude ($V_{peak}$) of $V_{load}$ for different $R_{load}$ and MW powers. Solid curves are best fit ones from theoretical calculations – see the main text.

FIG. 3: Load resistance ($R_{load}$) dependence of (a) electric current ($I_{load}$) and (b) dissipated power ($P_{dis}$) across the resistors, for different MW powers. Experimental data are represented by dot and solid curves are from bet fit results using theory – see the main text.

FIG. 4: Circuit diagrams representing spin-pumping/ISHE experiments carried out for this study, with the ISHE cell being modelled as (a) an non-ideal voltage, or (b) non-ideal current source.

**Fig. 1: Rogdakis et al.**

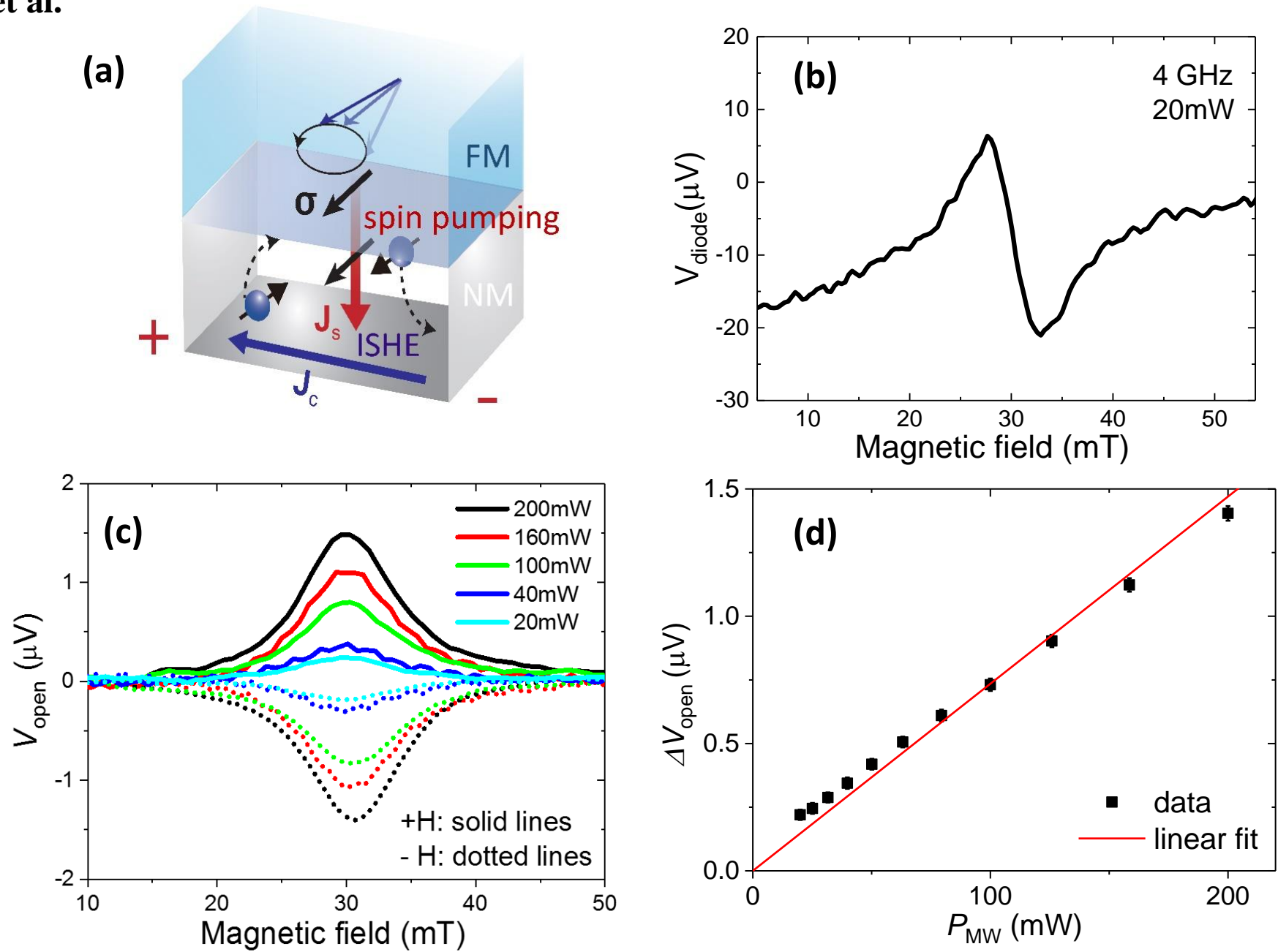

**Fig. 2: Rogdakis et al.**

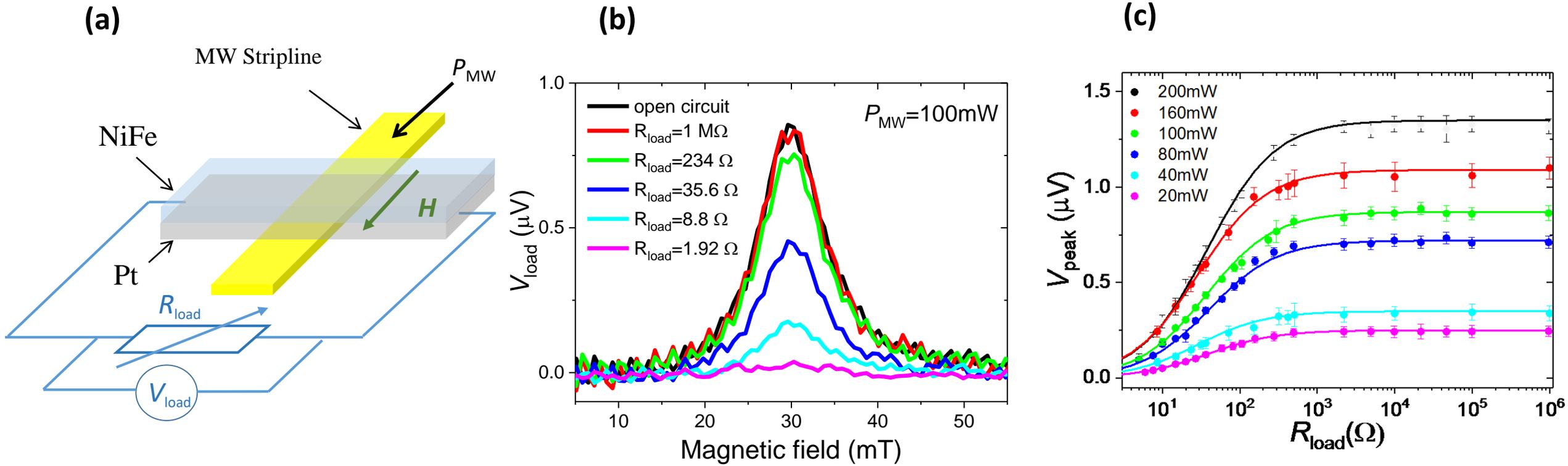

**Fig. 3: Rogdakis et al.**

(a) 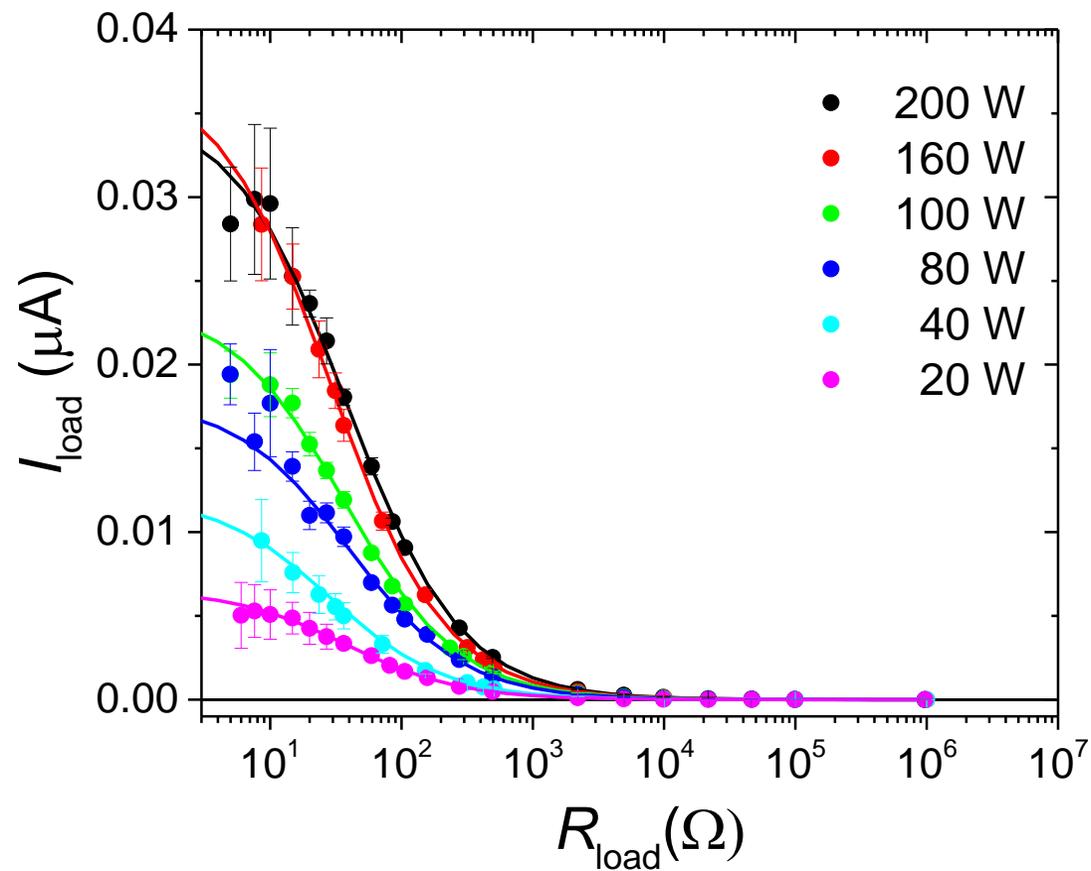

(b) 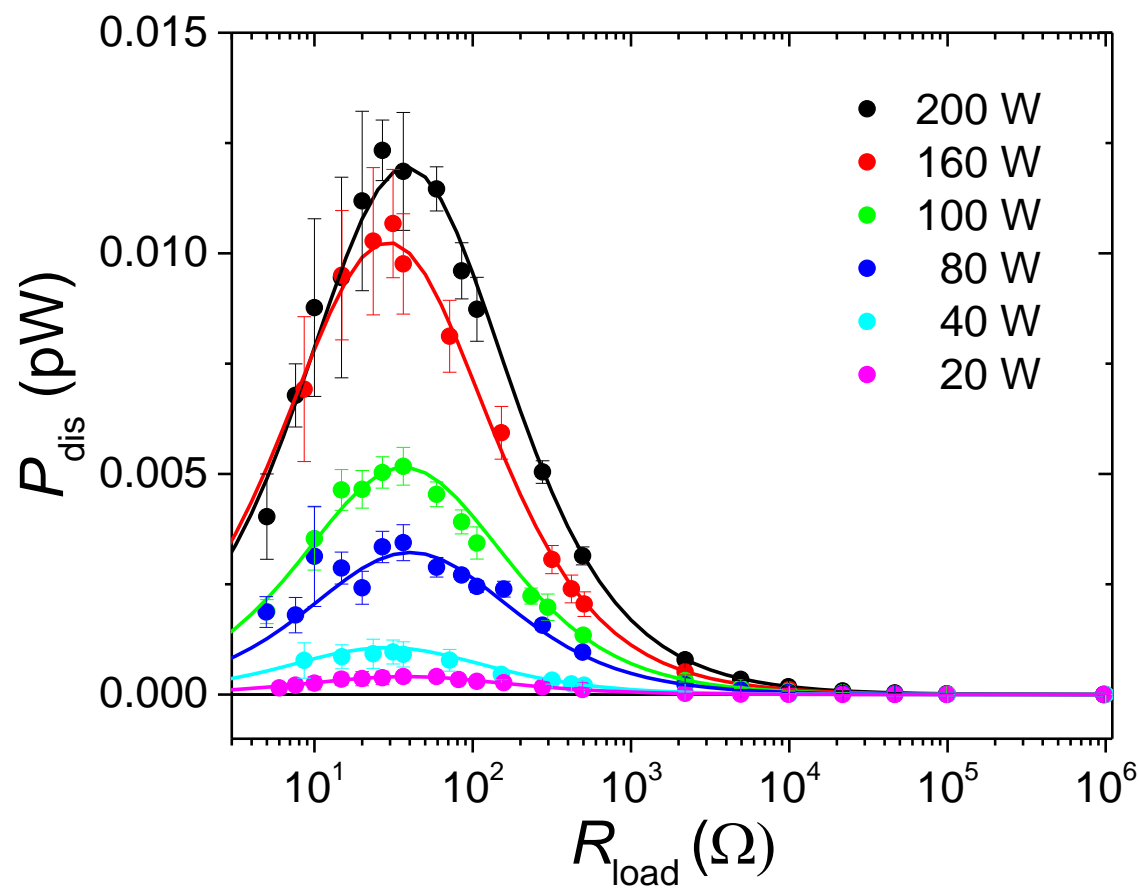

**Fig. 4: Rogdakis et al.**

(a) 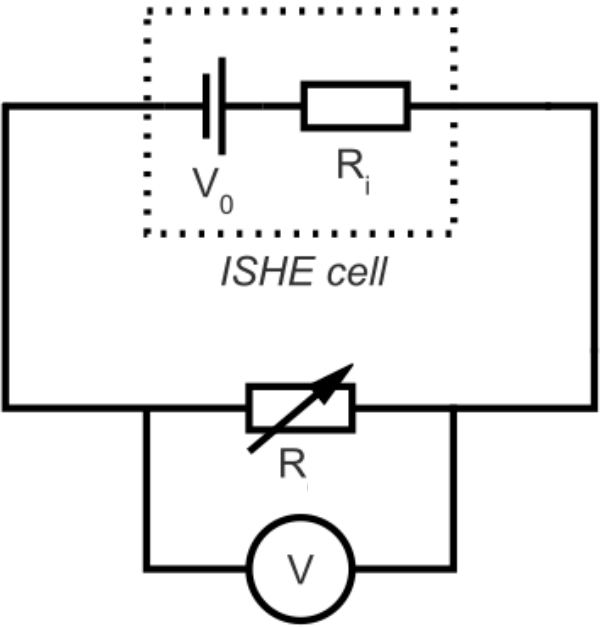

(b) 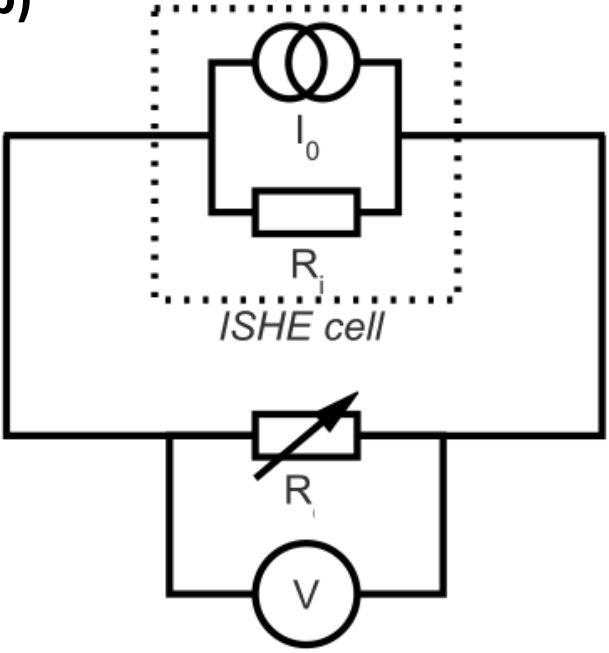